\documentclass[twocolumn,showpacs,preprintnumbers,amsmath,amssymb,prl,epsf]{revtex4}
\usepackage{graphicx}
\begin{document}
\title{Dispersion spreading of polarization-entangled states of light and two-photon interference}
\author{G. Brida, M. Genovese, L.~A.~Krivitsky}
\affiliation{Istituto Nazionale di Ricerca Metrologica,\\
Strada delle Cacce 91, 10135 Torino, Italy }
\author{M.~V.~Chekhova }
\affiliation{Department of Physics, M.V.Lomonosov Moscow State
University,\\  Leninskie Gory, 119992 Moscow, Russia} \vskip 24pt

\begin{abstract}
\begin{center}\parbox{14.5cm}
{We study the interference structure of the second-order intensity
correlation function for polarization-entangled two-photon light
obtained from type-II collinear frequency-degenerate spontaneous
parametric down-conversion (SPDC). The structure is visualised due
to the spreading of the two-photon amplitude as two-photon light
propagates through optical fibre with group-velocity dispersion
(GVD). Because of the spreading, polarization-entangled Bell states
can be obtained without any birefringence compensation at the output
of the nonlinear crystal; instead, proper time selection of the
intensity correlation function is required. A birefringent material
inserted at the output of the nonlinear crystal (either reducing the
initial o-e delay between the oppositely polarized twin photons or
increasing this delay) leads to a more complicated interference
structure of the correlation function.}
\end{center}
\end{abstract}
 \maketitle \narrowtext
\vspace{-10mm}
\section{I. Introduction}

Propagation of nonclassical light through optical fibres is a
common technique in quantum optics. Fibres are nowadays used to
produce nonclassical light ~\cite{kumar}, but even more often,
they are applied in quantum communication for transmitting
nonclassical light over large distances. Therefore, it is
important to study how propagation through fibres influences
nonclassical states of light. In particular, we have discovered
~\cite{brida} that propagation of two-photon light through fibres
allows one to observe polarization interference structure in the
shape of the second-order intensity correlation function. In this
paper we investigate this effect in more detail and discuss its
relation to other effects connected with the propagation of
two-photon light in media with group-velocity dispersion (GVD).

Two-photon interference is one of the basic effects observed for
two-photon light. One speaks of two-photon interference whenever the
coincidence counting rate for two detectors registering correlated
photons, for instance generated via Spontaneous Parametric Down
Conversion (SPDC) depends on some phase delay, which can be
introduced in many different ways. To vary the phase, one can use
polarization transformations ~\cite{kiess}, or introduce a space
delay ~\cite{sergienko}, or vary the selected frequency
~\cite{viciani} or angle ~\cite{burlakov}.

Closely related to two-photon interference is the anti-correlation
effect called Hong-Ou-Mandel 'dip' effect. It is observed when there
are two different ways for a photon pair to produce a coincidence:
both photons being reflected by a beamsplitter or both photons
passing through it. Cancellation of the corresponding probability
amplitudes causes the 'dip' in the coincidence counting rate. For
the interference to occur it is important that the two probability
amplitudes should be made indistinguishable by providing equal paths
for the two photons on their way to the beamsplitter. In the
polarization version of the 'dip' effect, this is achieved by
inserting a birefringent slab (an e-o delay) after the crystal
producing photon pairs.

In the present paper, two-photon interference is registered in the
shape of the second-order correlation function using a standard
setup for observing the 'dip' effect in its polarization version.
The two interfering probability amplitudes are made
indistinguishable not by introducing an e-o delay after the crystal
or by inserting narrow-band filters in front of the detectors, but
instead, by spreading the two-photon amplitudes in an optical fibre.

The paper is organised as follows. In Section II, we discuss the
effect of the propagation through a medium with group-velocity
dispersion on the correlation function of two-photon light. In
Section III, we describe theoretically the interference structure
appearing in the shape of the second-order correlation function as
two-photon light generated via type-II SPDC propagates through an
optical fibre and is registered by a setup for observing the
anti-correlation effect. Sections IV, V are devoted to the
experimental observation of this effect: the experimental setup is
described in Section IV, and the results are presented in Section V.
Conclusions are made in Section VI.

\section{II. Spreading of the two-photon amplitude in an optical fibre}

In the general case, the state vector of two-photon light generated
via SPDC or four-wave mixing ~\cite{kumar} can be represented as

\begin{eqnarray}
|\Psi\rangle=|vac\rangle+\int{dk_1}{dk_2}F(k_1,k_2)a^{\dagger}_{k_1}
a^{\dagger}_{k_2}|vac\rangle, \label{1}
\end{eqnarray}

where $k_1,k_2$  are the modes into which the two photons are
emitted, $a^{\dagger}_{k_1}$ and  $a^{\dagger}_{k_2}$ are photon
creation operators in these modes, and  $F(k_1,k_2)$ is the
two-photon spectral amplitude. In the case of a continuous-wave
pump. $F(k_1,k_2)\propto\delta(\omega_1-\omega_2)$ , where
$\omega_1,\omega_2$ are the frequencies of the two photons. Let us
assume, for simplicity, that only photons emitted in certain
directions $k_1,k_2$ are selected ~\cite{distance}. Then
$F(k_1,k_2)$ contains only frequency dependence, and hence, the
state vector of two-photon light has the form

\begin{eqnarray}
|\Psi\rangle=|vac\rangle+\int{d\Omega}F(\Omega)a^{\dagger}_{1}(\omega_0+\Omega)
a^{\dagger}_{2}(\omega_0-\Omega)|vac\rangle, \label{2}
\end{eqnarray}

with $\omega_0$ being the central frequency of the spectrum of the
two-photon light, $\omega_1\equiv\omega_0+\Omega,
\omega_2\equiv\omega_0-\Omega$ .

The spectral and temporal properties of two-photon light are fully
described by the two-photon spectral amplitude  $F(\Omega)$. It was
shown in ~\cite{chekhova} that the first-order and second-order
correlation functions of two-photon light are given by the following
expressions:

 \begin{eqnarray}
 G^{(1)}(\tau)\propto\int{d\Omega}|F(\Omega)|^2\exp(i\Omega\tau)\nonumber
 \\
  G^{(2)}(\tau)\propto|\int{d\Omega}F(\Omega)\exp(i\Omega\tau)|^2
 .\end{eqnarray}

The Fourier transform of the two-photon spectral amplitude,
$F(\tau)\equiv\int{d\Omega}F(\Omega)\exp(i\Omega\tau)$  is usually
called the two-photon amplitude.

Although the first-order and the second-order correlation functions
are both determined by the two-photon spectral amplitude, there is
an important difference between them. While the first-order
correlation function contains no information about the phase of
$F(\Omega)$ , the second-order correlation function essentially
depends on this phase. This is why the second-order correlation
function spreads when two-photon light propagates through a medium
with group-velocity dispersion (GVD). Indeed, as shown in
~\cite{chekhova} and ~\cite{valencia}, the behavior of the
two-photon amplitude in a GVD medium is similar to the behavior of a
short optical pulse. After propagation through a GVD medium of
length $z$, the two-photon amplitude $F(\tau)$ turns into

\begin{equation}
\widetilde{F}(\tau)=\frac{1}{\sqrt{4\pi i k''
z}}e^\frac{i(\tau-k'z)^2}{4k''z}F(\Omega)|_{\Omega=\frac{\tau-k'z}{2k''z}},
 \label{spread}
\end{equation}

where $k',k''$ are, respectively, the first and the second
derivatives of the fibre dispersion law $k(\omega)$  for the
transmitted photons. If these parameters are different for the two
photons of the pair, their mean arithmetic values enter the formula.

In experiment, one does not observe the two-photon amplitude, but
its square module, the second-order correlation function. The effect
of $G^{(2)}$ spreading was indeed observed by passing two-photon
light through a 500m optical fibre and measuring the coincidence
time distribution at the output of the fibre ~\cite{valencia}. The
width of $G^{(2)}$, initially on the order of $100$ fs, grew up to
several nanoseconds, which enabled its experimental observation.
Here it is worth mentioning that the 'spreading' effect is so far
the only way to determine the initial width of the second-order
correlation function. Indeed, since the final width of $G^{(2)}$ is,
like in the case of a short pulse, related to its initial width, the
initial width can be found by measuring the final width. This is
important, since because of the extremely small initial width of
$G^{(2)}$, there are no other experimental techniques for measuring
it.

It is sometimes claimed that the width and shape of the
second-order correlation function can be found by measuring the
width and shape of the anti-correlation 'dip'. However, the shape
of the 'dip' corresponds to $G^{(1)}(\tau)$ rather than
$G^{(2)}(\tau)$ ~\cite{karabutova} and hence does not contain any
extra information with respect to the spectrum. The fact that the
'dip' shape is in one-to-one correspondence with the spectrum
explains in the most simple way the 'dispersion cancellation
effect' ~\cite{steinberg}, i.e., the fact that the 'dip' shape
remains the same if one or both photons propagate through a
dispersive medium. Indeed, it is $G^{(2)}(\tau)$ and not
$G^{(1)}(\tau)$ that is spread in a dispersive medium and, since
both the spectrum and the first-order correlation function are
insensitive to GVD, the shape of the 'dip' remains unchanged in
the presence of GVD.

Thus, similar to the way a short pulse gets spread in a GVD medium
and at a sufficiently large distance acquires the shape of its
spectrum, the two-photon amplitude after a sufficiently long GVD
medium acquires the shape of its Fourier transform, the two-photon
spectral amplitude, i.e. a GVD medium performs the Fourier
transformation of the two-photon amplitude. This fact can be used to
reveal the interference structure contained in the two-photon
spectral amplitude of light in an experiment on observing the 'dip'
effect in its polarization version.

\section{III. Interference structure of the spread intensity correlation function}

When the anti-correlation 'dip' is observed in its polarization
version, the experiment is organized as follows (Fig.1). The
two-photon light emitted via type-II SPDC under collinear
frequency-degenerate phase matching is split in two beams by a
non-polarizing beamsplitter, and each of the two beams, after
passing through a polarizer, is addressed to a photon counting
detector. The output pulses of both detectors are sent to a
coincidence circuit, and the coincidence counting rate is measured.
There are two basic ways to observe the interference. In one way,
the polarizers are fixed at the positions ($45^\circ, -45^\circ$) or
($45^\circ, 45^\circ$) and a variable delay is introduced between
the photons of a pair. When the delay exactly equals $DL/2$, where
$L$ is the crystal length and
$D\equiv\frac{1}{u_{o}}-\frac{1}{u_{e}}$ is the inverse group
velocity difference between the orthogonally polarized photons of a
pair, the coincidence counting rate goes to a minimum or to a
maximum, depending on the polarizer settings. The other way to
observe the 'dip' is to fix the delay and to rotate one of the
polarizers, the other one being fixed at $45^\circ$. This way one
obtains an interference pattern with the visibility depending on the
e-o delay. When the e-o delay equals $DL/2$, the visibility is
maximal and close to $100\%$. This pattern is usually called
polarization interference pattern.

The fact that obtaining high-visibility interference requires the
e-o delay to be equal to $DL/2$ follows from the following
considerations. In a setup as shown in Fig.1, the state at the
output of the beamsplitter is
\begin{eqnarray}
|\Psi\rangle=|vac\rangle+\int{d\Omega}F(\Omega)
\{a^{\dagger}_{H1}(\omega_0+\Omega)
a^{\dagger}_{V2}(\omega_0-\Omega)e^{i\Omega\tau_0} \nonumber
\\
+a^{\dagger}_{V1}(\omega_0+\Omega)
a^{\dagger}_{H2}(\omega_0-\Omega)\exp^{-i\Omega\tau_0}\}|vac\rangle,
\label{5}
\end{eqnarray}

where $a^{\dagger}_{\sigma i}$ are photon creation operators in the
horizontal and vertical polarization modes (denoted by $\sigma=H,V$)
and two spatial modes (denoted by $i=1,2$); $\omega_0=\omega_p/2$,
$\omega_p$ is the pump frequency, and the terms corresponding to
both photons of the two-photon state going to the same port of the
beamsplitter have been omitted. The difference between the group
velocities of signal and idler photons in the non-linear crystal
where spontaneous parametric down-conversion is generated, leads to
the factors $e^{\pm i \Omega\tau_0}$ by the two terms of Eq.(1),
with $\tau_0=DL/2$.

If one polarizer is fixed at an angle $\theta_1$  and the other
polarizer, at an angle $\theta_2$, then, taking into account the
transformation of the photon creation operators, we can write the
state after the polarizers in the same way as in Eq.(2), but with
the spectral two-photon amplitude $F(\Omega)$ replaced by

\begin{eqnarray}
F'(\Omega)=F(\Omega)\{\sin(\theta_1+\theta_2)\cos(\Omega\tau_0)\nonumber
\\-i\sin(\theta_1-\theta_2)\sin(\Omega\tau_0)\}.
 \label{6}
\end{eqnarray}

The second-order correlation function is given by the square modulus
of the two-photon amplitude , which can be easily calculated from
Eq.(6):

\begin{eqnarray}
F'(\tau)=\cos(\theta_1)\sin(\theta_2)F(\tau+\tau_0) \nonumber
\\+\cos(\theta_2)\sin(\theta_1)F(\tau-\tau_0).
 \label{7}
\end{eqnarray}

Since the two-photon spectral amplitude for type-II SPDC has the
shape
\begin{equation}
 F(\Omega)=\frac{\sin(DL\Omega/2)}{DL\Omega/2},
 \label{8}
\end{equation}
the two-photon amplitude  $F(\tau)$ has a rectangular shape with the
width  $2\tau_0$ ~\cite{Rubin}. For this reason, the first and the
second terms of Eq.(7) do not overlap in time, and no interference
can be observed between them in the second-order correlation
function. However, it is worth mentioning that the two-photon
amplitude $F'(\tau)$ indeed has interference structure, and it could
be revealed if one were able to measure $F'(\tau)$ directly and not
its square modulus (the intensity correlation function).

The two cases demonstrating interference most explicitly are the
following ones~\cite{kiess}: $\theta_1=\theta_2=45^\circ$  and
$\theta_1=-45^\circ, \theta_2=45^\circ$. For these cases, Eq.(7)
gives the following second-order intensity correlation functions:

 \begin{equation}
 G_{\pm}^{(2)}(\tau)\equiv|F'(\tau)_\pm|^2=\frac{1}{4}|F(\tau-\tau_0)\pm
 F(\tau+\tau_0)|^2,
 \label{9}
\end{equation}
where $G_{+}^{(2)}(\tau)$ corresponds to
$\theta_1=\theta_2=45^\circ$ and $G_{-}^{(2)}(\tau)$ to
$\theta_1=-45^\circ, \theta_2=45^\circ$ orientations of the
polarizers.

To observe the interference, one usually compensates the delay
$\tau_0=DL/2$ by introducing additional birefringent material after
the crystal ~\cite{Rubin},~\cite{delay}. When $\tau_0$ is reduced to
zero, the two-photon amplitude in Eq.(7) becomes
$F'(\tau)=\sin(\theta_1+\theta_2)F(\tau)$ , and the interference
with 100\% visibility can be observed by fixing one of the
polarizers at and rotating the other one.

Apart from compensating the delay, interference can be obtained by
using narrowband filters selecting the central part of the spectral
amplitude $F(\Omega)$ ~\cite{Rubin}. This can be viewed as
'spreading' the two amplitudes in Eq.(9), to provide an overlap
between them.

In the present work, we spread the two-photon amplitudes in Eq.(9)
in a different way: not by using narrowband filters but by passing
the two-photon radiation through a sufficiently long optical fibre.
This, on the one hand, provides the same effect as narrowband
filters, in the sense that high-visibility polarization interference
can be observed. On the other hand, it allows one to observe the
interference structure in the shape of the second-order correlation
function, which, due to the propagation of two-photon light through
the fibre, becomes very broad.

Indeed, if the biphoton beam, before entering the beamsplitter,
passes through a fibre with length $z$ being large enough
~\cite{sufflength}, then, according to (\ref{spread},9)

\begin{eqnarray}
G_{\pm}^{(2)}(\tau')\sim|e^{\frac{i(\tau'-\tau_0)^2}{2\tau_0\tau_f}}\hbox{sinc}(\frac{\tau'-\tau_0}{\tau_f})
\nonumber\\
\pm
e^{\frac{i(\tau'+\tau_0)^2}{2\tau_0\tau_f}}\hbox{sinc}(\frac{\tau'+\tau_0}{\tau_f})|^2,
 \label{pmspread}
\end{eqnarray}
where $\tau'\equiv\tau-k'z$ is the shifted time and
$\tau_f\equiv2k''z/\tau_0$  is the typical width of the correlation
function after the fibre ~\cite{Krivitsky}.

Taking into account that $\tau_f>>\tau_0$ , one can rewrite
Eq.(\ref{pmspread}) as
\begin{eqnarray}
G_{+}^{(2)}(\tau')\sim\frac{\sin^2(\tau'/\tau_f)\cos^2(\tau'/\tau_f)}{(\tau'/\tau_f)^2},\nonumber\\
G_{-}^{(2)}(\tau')\sim\frac{\sin^4(\tau'/\tau_f)}{(\tau'/\tau_f)^2}.
\label{sinecosine}
\end{eqnarray}

The interference structure of the correlation functions
Eq.(\ref{sinecosine}) can be observed if the time resolution of
measuring $G^{(2)}(\tau')$ is good enough. Two main reasons for the
resolution reduction must be considered: the time jitter of the
detectors (typically of the order of several hundreds of
picoseconds), and the jitter contribution of the electronic
technique (amplitude walk and noise) used for the coincidence
detection.

The most widely used coincidence detection technique involves direct
measurement of the time delay between the photocount pulses of the
two detectors by means of a time-to-amplitude converter (TAC), which
converts linearly the time interval between the two input pulses
(START and STOP) into an output pulse of a proportional amplitude.
This analog pulse is forwarded to a multichannel analyzer (MCA),
which gives the histogram of the input pulse amplitudes
corresponding to the probability distribution for the time interval
between the counts of the two detectors. One can show ~\cite{Mandel}
that in the limit of small photon fluxes, this time interval
distribution coincides in shape with the second-order intensity
correlation function. The resolution of such technique could be on
the order of one picosecond if the detector jitter time were
negligible. Therefore, it is mainly the time jitter of the detectors
that sets the resolution of measuring the shape of $G^{(2)}(\tau')$.
It follows that in order to observe the interference structure, the
time spread of the correlation function in the fibre should be much
larger than the jitter time of the detectors.

It is interesting to discuss how the correlation function changes if
the delay $\tau_0$ between the signal and idler photons is changed
by placing after the crystal a birefringent plate (for instance, a
quartz plate) introducing a delay $\tau_p$  with the sign equal or
opposite to that of $\tau_0$ . Formulas (\ref{sinecosine}) in this
case become
\begin{eqnarray}
\nonumber
G_{+}^{(2)}(\tau')\sim\frac{\sin^2(\tau'/\tau_f)\cos^2(\kappa\tau'/\tau_f)}{(\tau'/\tau_f)^2},\nonumber\\
G_{-}^{(2)}(\tau')\sim\frac{\sin^2(\tau'/\tau_f)\sin^2(\kappa\tau'/\tau_f)}{(\tau'/\tau_f)^2},
\label{plates}
\end{eqnarray}

where $\kappa=\frac{\tau_{0} + \tau{p}}{\tau_{0}}$ . For instance,
if $\tau_{p} =-1/2 \tau_0$ (the plate compensates only for half of
the e-o delay in the crystal ), $\kappa=\frac{1}{2}$ and the
modulation period becomes twice larger than in the absence of the
plate.

If the delay $\tau_0$ is compensated completely, the
$G^{(2)}_{-}(\tau')$ peak disappears while the peak
$G^{(2)}_{+}(\tau')$ acquires the same shape as the peak observed
without the polarizers.

If the sign of the delay $\tau_{p}$ is the same as that of $\tau_0$,
the modulation period of the correlation function after the fibre
becomes smaller than in the absence of the birefringent plate.

The next two sections are devoted to the experimental observation of
these effects.

\section{IV. Experimental setup}

Two-photon light was generated via spontaneous parametric
down-conversion by pumping a type-II 0.5 mm $\beta$-barium borate
 crystal (BBO) with 0.2 Watt $\hbox{Ar}^{+}$ cw laser beam at the
wavelength 351 nm in the collinear frequency-degenerate regime
(Fig.2). After the crystal, the pump laser beam was eliminated by a
95\% reflecting UV mirror and the SPDC radiation was coupled into a
single-mode fibre with the length 240 m or 1 km and the mode field
diameter (MFD) of $4\mu$m by a 20x microscope objective lens placed
at the distance of 50 cm. This scheme provided imaging of the pump
beam waist onto the fibre input with the magnification 1:40. Then,
the angle selected by the numerical aperture (NA) of the fibre from
the SPDC angular spectrum was $\delta\Theta\approx NA/40=0.006$rad.
Without this narrow angular selection, the interference structure
would be smeared by contributions from various parts of the SPDC
angular spectrum. For the efficient use of the pump beam, the pump
was focused into the crystal using a UV-lens with the focal length
30 cm. The resulting beam waist diameter, $120\mu$m, was small
enough to be coupled with the fibre core diameter but still
sufficiently large not to influence the SPDC angular spectrum.

Because of a rapid polarization drift in the fibre, a quarter-wave
plate (QWP) and a half-wave plate (HWP) were introduced after the
output of the fibre. These two plates were adjusted to compensate
for the random polarization transformation in the fibre and recover
the initial horizontal- vertical polarization basis. After the
fibre, the biphoton pairs were addressed to a 50/50 nonpolarizing
beamsplitter and two photodetection apparatuses, consisting of
red-glass filters, pinholes, focusing lenses and avalanche
photodiodes (SPCM by EG\&G). The photocount pulses of the two
detectors, after passing through delay lines, were sent to the START
and STOP inputs of a TAC. The output of the TAC was finally
addressed to an MCA, and the distribution of coincidences over the
time interval between the photocounts of the two detectors,
$R_c(\tau')$, was observed at the MCA output.

The FWHM of the coincidence peak in the absence of the fibre was
measured to be approximately 0.8 ns, a value substantially
determined by the APD time jitter. In the presence of the fibre, the
FWHM of the peak increased, to $1.2$ ns in the case of the $240$ m
fibre (Fig.3a) and to $4.5$ ns in the case of the $1000$ m fibre
(Fig.3b). The observed distributions were fitted using
Eq.(\ref{spread}) with the following fitting parameters: position of
the peak, its height, the GVD of the fibre, and the background level
caused by accidental coincidences. The obtained GVD values for both
fibres used were $3,2\times10^{-28}$ s$^2$/cm.

The distributions of Fig.3 were obtained without any polarization
selection after the beamsplitter. In the main part of the
experiment, two polarizers (Glan prisms) were inserted in front of
the detectors, and the coincidence distribution was analyzed for the
$\theta_1=\theta_2=45^\circ$  and $\theta_1=-45^\circ,
\theta_2=45^\circ$ settings of the polarizers.

\section{V. Results and discussion}

To observe polarization interference in the shape of the correlation
function after the fibre, the coincidence distribution at the MCA
output was recorded for the two settings of the polarizers
considered above, $\theta_1=\theta_2=45^\circ$  and
$\theta_1=-45^\circ, \theta_2=45^\circ$. This measurement was made
for two different fibres: a $240$ m one (Fig. 4) and a $1$ km one
(Fig. 5). In the case of the $240$ m fibre, the spread of the
correlation function was insufficient for providing
 100\% interference visibility. Indeed, for the width of the correlation function $1.2$ ns
 and the time resolution of the measurement (determined by the detectors time jitter) $750$ ns,
 the theory predicts a visibility of $48\%$. (The visibility of polarization interference
 is obtained from the maximum value in the center of the correlation
 function in the case of constructive interference ($\theta_1=\theta_2=45^\circ$) and corresponding minimum value in the case of destructive
 interference ($\theta_1=-45^\circ, \theta_2=45^\circ$). The experiment
 showed even less visibility (about 35\%), which was probably caused by the polarization drift in
 the fibre. (Although polarization state of the two-photon light after the fibre was measured and
 corrected every 10 minutes, polarization drift between these control moments
 justifies the decrease in the interference visibility from $48\%$ to $35\%$.)

With a $1$ km fibre, the spread of the correlation function ($4.5$
ns) is large enough, so that the interference structure is not
significantly smeared by the detectors time jitter (Fig.5). As the
settings of the polarization prisms change from
$\theta_1=\theta_2=45^\circ$ to $\theta_1=-45^\circ,
\theta_2=45^\circ$, the coincidence counting rate at the center of
the correlation function distribution decreases 5.4 times. It means
that with a proper selection of the coincidence time window, one can
observe high-visibility polarization interference. Indeed, Fig.6
shows the dependence of coincidence counting rate on the orientation
of the first polarizer, $\theta_1$, with the other polarizer fixed
at $\theta_2=45^\circ$ and only coincidences from a time window of
$0.43$ ns (7 MCA channels) near the center of the peak being
registered. The visibility of the observed polarization interference
fringes is $78\%$. This dependence indicates that by spreading the
second-order intensity correlation function and time-selecting it, a
Bell state can be produced. To test this we calculated the value of
violation of Bell inequality
$R\equiv(N(\theta)-N(3\theta))/N(\infty)\leq0.25$, where $N(\theta)$
is the number of coincidences taken at the relative orientation of
polarizers $\theta\equiv\theta_1-\theta_2$ which have been chosen
$\theta=30^\circ$ and $N(\infty)$ is the number of coincidences
measured without any polarization selection ~\cite{GenovesePhysRep}.
After subtracting the accidental coincidences we observed a
 violation of Bell inequality $R=0.322\pm0.061$.

As a method of producing a Bell state through SPDC in a type-II
crystal, the technique used here is very similar to using
narrow-band filters: instead of frequency-selecting the spectral
amplitude, here we first Fourier-transform it and then perform time
selection. However, if narrow-band filters are used for selecting a
Bell state, they are usually centered at the degenerate SPDC
frequency. In contrast, in our case, time selection does not have to
be performed at the zero time delay. From Eqs.(11) Fig.5, it is
clear that high-visibility interference can be also observed at the
slopes of the peak or at its side lobes.

For future comparison of the experiment with theoretical
predictions, we also measured the correlation function distributions
in the case where a birefringent material (quartz) was introduced
after the crystal. We used a single quartz plate, with thickness $1$
mm, whose optic axis could be set parallel to the plane containing
the optic axis of the BBO crystal or orthogonal to it. In the first
case, the quartz plate reduced the e-o delay between the photons of
the same pair while in the second case, it increased this delay. The
corresponding coincidence distributions obtained in experiment are
presented in Figs.7 (a-d). The curves show the theoretical fit with
formulas~(\ref{plates}).

In all measurements described above, the main difficulty was caused
by the rapid polarization drift in the fibre, which considerably
smeared the interference structure of the second-order correlation
function. To avoid this effect, we have performed a measurement
where polarization drift was eliminated, by placing a single
polarization prism before the fibre and no prisms after the fibre.
Naturally, this way one can only observe the coincidence
distribution in the $\theta_1=\theta_2=45^\circ$ configuration,
since both polarizers are represented by a single Glan prism fixed
at $45^\circ$. Coincidence distributions obtained this way are shown
in Fig.8: (a) with no plates; (b) with one $1$ mm plate and (c) with
two $1$ mm plates inserted after the crystal, their optic axes being
orthogonal to the plane of the BBO crystal optic axis. Here it is
clearly seen that increasing the e-o delay (by inserting additional
plates) leads to a decrease in the oscillation period in the shape
of the second-order correlation function. The interference structure
observed in Fig.8c has relatively low visibility because the
modulation period becomes comparable with the time resolution of the
detection system ($750$ ns).

\section{VI. Conclusion}

In summary, we have studied, both theoretically and experimentally,
the polarization interference structure of the second-order
intensity correlation function for two-photon light generated via
type-II SPDC and fed into a setup for observing the anti-correlation
'dip' effect. To register this structure, we made the correlation
function get spread due to the propagation of the two-photon light
through a medium with group-velocity dispersion (in our case, an
optical fibre). The interference reveals itself as a peak or a dip
in the middle of the coincidence distribution observed at the MCA
output. Since the interference structure is caused by the delay
between orthogonally polarized photons of a pair accumulated in the
course of their propagation through the crystal, increasing or
reducing this delay by means of birefringent plates introduced after
the crystal considerably changes the structure. This effect has been
also observed in the experiment.

The shape of the correlation function after the fibre is given by a
scaled square modulus of the two-photon spectral amplitude. One can
say that the fibre performs a Fourier transformation of the
two-photon amplitude. This fact, already mentioned earlier in the
literature ~\cite{valencia}, can be also used for measuring the
width of the two-photon amplitude in the cases where it is too
narrow to be measured by alternative methods.

It might seem that because the shape of the two-photon amplitude
after the fibre repeats, with some scaling factor, the two-photon
spectral amplitude before the fiber, the spectrum of two-photon
light should reveal the same structure. This, however, is not true,
because in our consideration of the state vector (see Eq.(5)) we
only included the terms corresponding to the cases where both
photons went into different ports of the beamsplitter. At the same
time, the spectrum of SPDC is formed by all kinds of pairs,
including those going into a single port.

Still, by fixing a time delay in the coincidence distribution after
the fibre we effectively fix the frequency offset $\Omega$  for one
of the photons of a pair (and for the other one as well). Does it
mean that the presented way of observing polarization interference
is identical to the one where narrow-band filters are used,
selecting the frequencies of the two photons? In principle, the
answer is 'yes'. However, in practice, with the time delay selection
one can use not only the central part of the coincidence
distribution but also its side parts, which is much more difficult
to provide with the filters.

It is important to stress that the interference would be present in
the two-photon amplitude and hence, in the two-photon spectral
amplitude, after the NPBS and the Glan prisms, even without the
fibre. However, without the fibre, all that one can measure is the
square modulus of the two-photon amplitude, which has no
interference structure. Measuring the shape of the two-photon
spectral amplitude after the beamsplitter is difficult since it
requires spectroscopic and correlation techniques simultaneously. In
our work it is done with the help of the fibre: the spread
two-photon amplitude reproduces the shape of the spectral amplitude,
and is broad enough to be studied with the existing equipment.

Finally, one can mention that the experiment presented here can be
considered as a method of preparing (with post-selection)
polarization Bell states in the two space modes after the
beamsplitter. This fact by itself has little practical importance
since Bell-state preparation with two-photon light is a
well-developed field of quantum optics and there are tens of various
ways to generate Bell states. However, it is worth noting that
effects of  $G^{(2)}$ spreading will take place whenever two-photon
light is propagating through a sufficiently long fibre (which is the
case in many quantum communication experiments), and this can be
taken into account to simplify the observation of two-photon
interference.

This work has been supported by MIUR (FIRB RBAU01L5AZ-002 and
RBAU014CLC-002, PRIN 2005023443-002 ), by Regione Piemonte (E14),
and by "San Paolo foundation", M.Ch. also acknowledges the
support of the Russian Foundation for Basic Research, grant
no.06-02-16393.

\begin{figure}
\includegraphics[height=4cm]{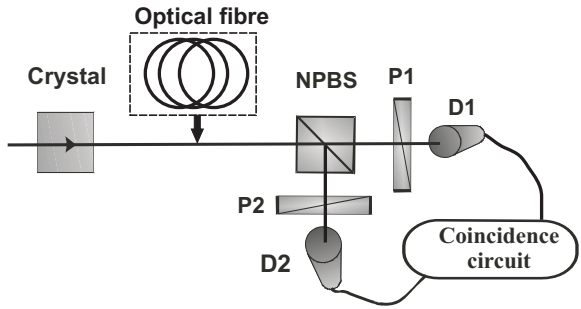} \caption{Typical setup for
observing two-photon polarization interference for type-II SPDC.
NPBS-50/50 nonpolarizing beamsplitter; P1 and P2- linear
polarization filters; D1, D2- photodetectors}
\end{figure}

\begin{figure}
\includegraphics[height=5cm]{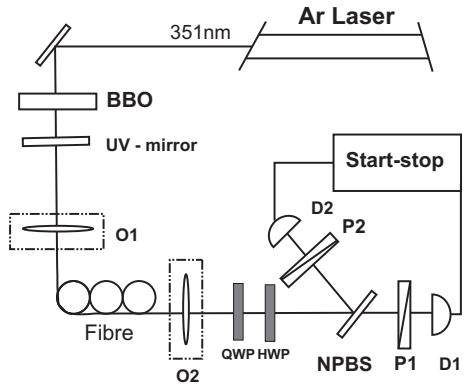} \caption{The experimental
setup. A cw $Ar^+$ laser at 351 nm pumps a type-II BBO crystal; O1,
O2- microscope objectives; QWP and HWP- retardation plates; NPBS -
50/50 nonpolarizing beamsplitter; P1 and P2- Glan prisms; D1, D2-
avalanche photodiodes. The output of the Start-Stop scheme is
analyzed by a multi channel analyzer (MCA).}
\end{figure}

\begin{figure}
\includegraphics[height=5cm]{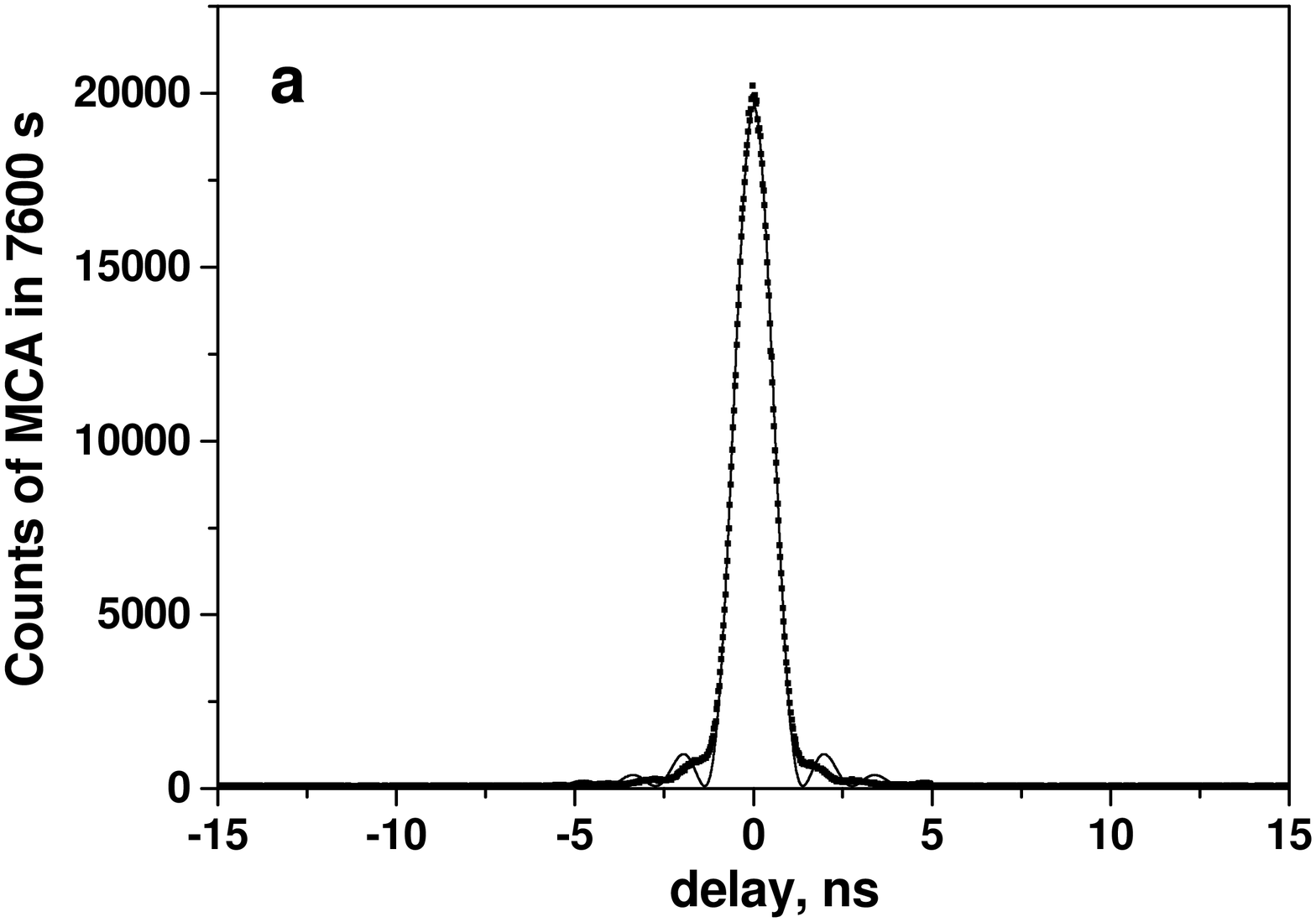}
\includegraphics[height=5cm]{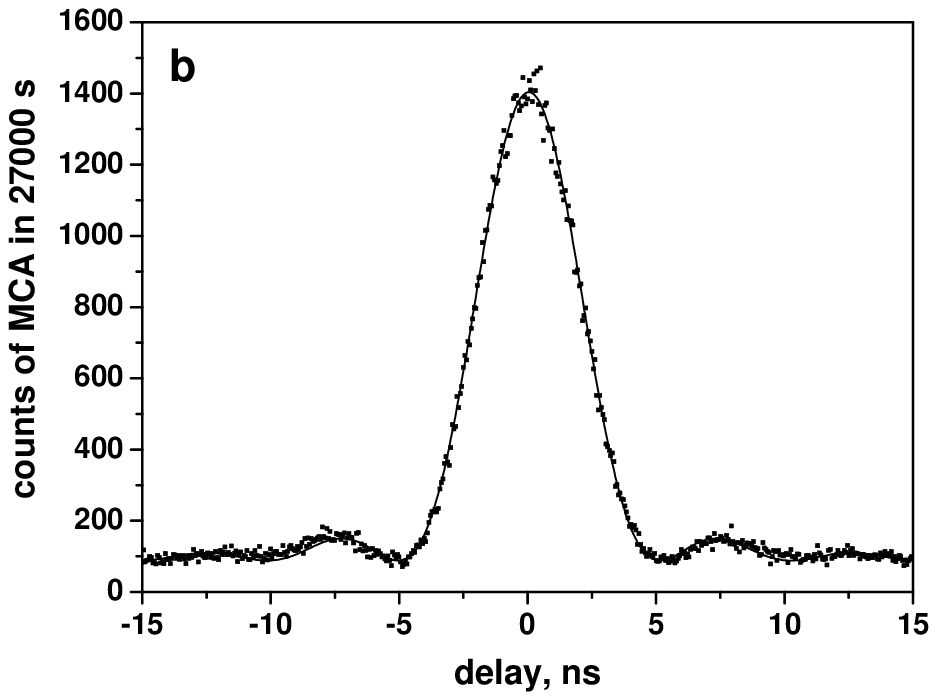}
 \caption{The second-order intensity correlation function of two-photon light
after (a) a 240-m fibre and (b) a 1000-m fibre.}
\end{figure}

\begin{figure}
\includegraphics[height=5cm]{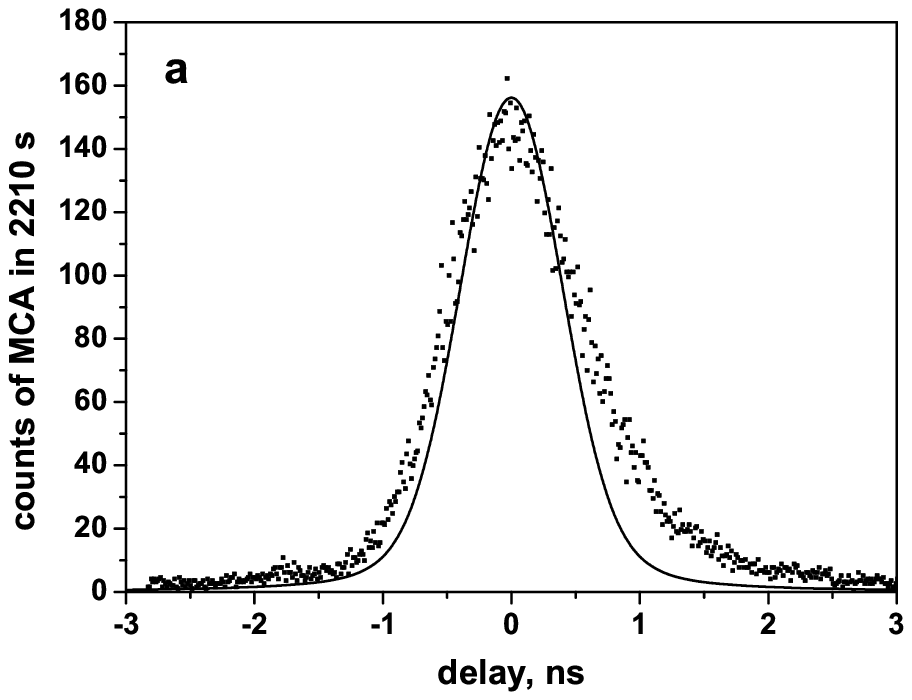}
\includegraphics[height=5cm]{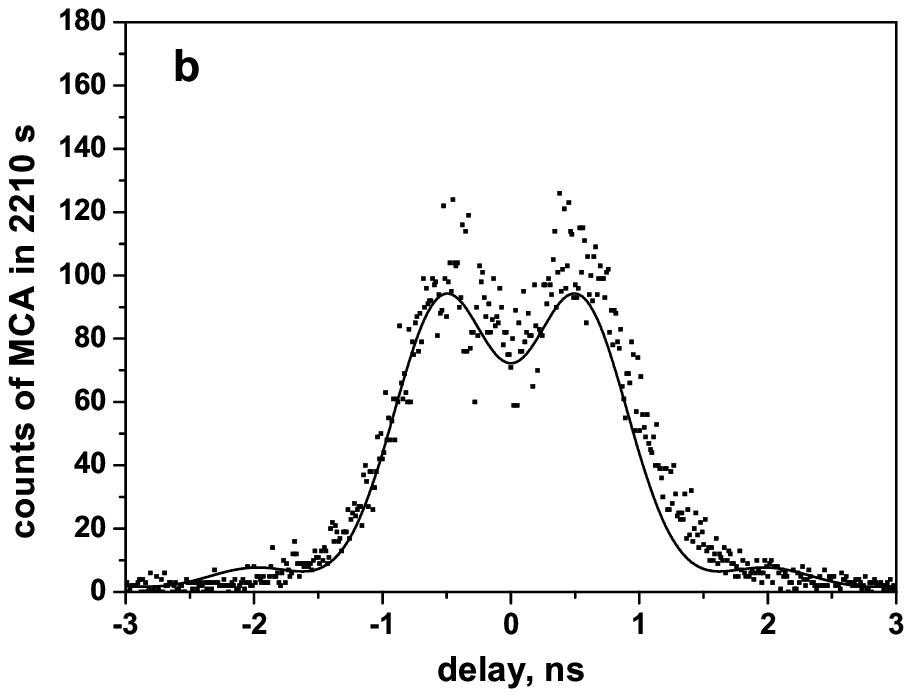}
 \caption{The measured shape of the second-order correlation function in the
case of a 240-m fibre and the settings of the polarizers (a)
$\theta_1=\theta_2=45^\circ$ and (b) $\theta_1=-45^\circ,
\theta_2=45^\circ$ . Points are experimental data from the MCA
output; curves show the theoretical dependencies with an account for
the detector jitter time $750$ ns.}
\end{figure}

\begin{figure}
\includegraphics[height=5cm]{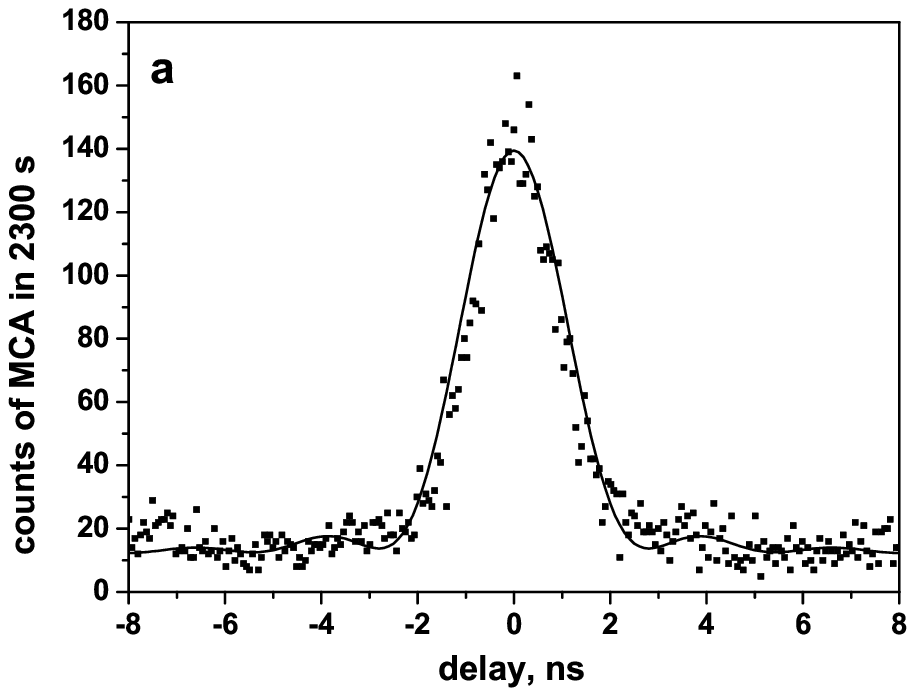}
\includegraphics[height=5cm]{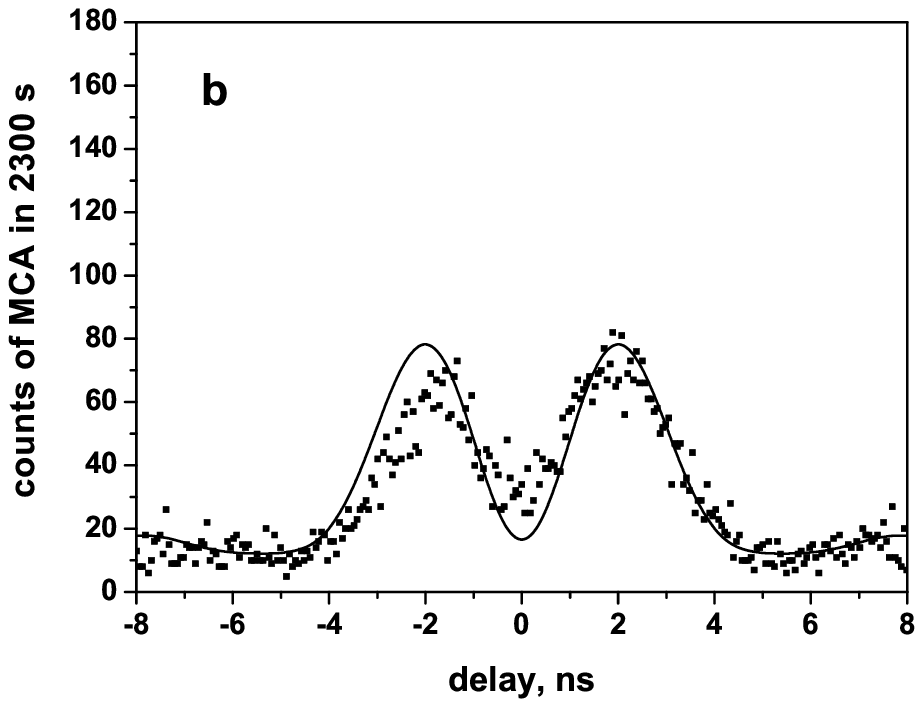}
 \caption{The measured shape of the second-order correlation function in the
case of a 1-km fibre and the settings of the polarizers (a)
$\theta_1=\theta_2=45^\circ$ and (b) $\theta_1=-45^\circ,
\theta_2=45^\circ$ . Points are experimental data from the MCA
output; curves show the theoretical dependencies with an account for
the detector jitter time $750$ ns.}
\end{figure}

\begin{figure}
\includegraphics[height=5cm]{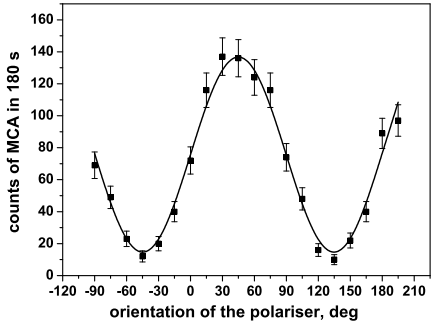} \caption{Coincidence counting rate as a function of the orientation of
polarizer 1, with polarizer 2 fixed at $45^\circ$  and only
coincidences within a 0.43 ns window at the center of the peak being
registered.}
\end{figure}

\begin{figure}
\includegraphics[height=5cm]{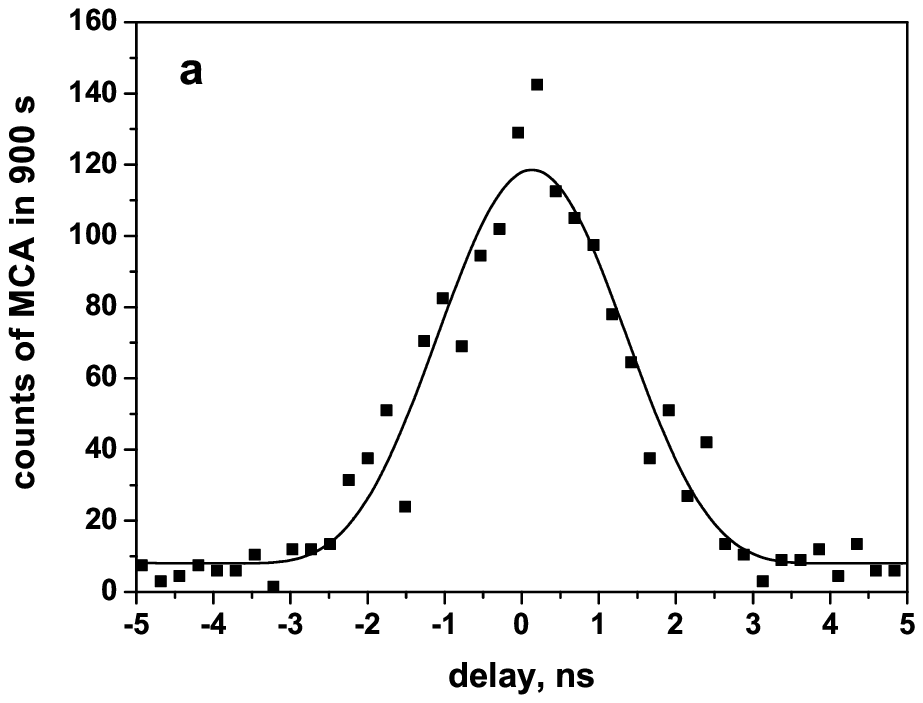}
\includegraphics[height=5cm]{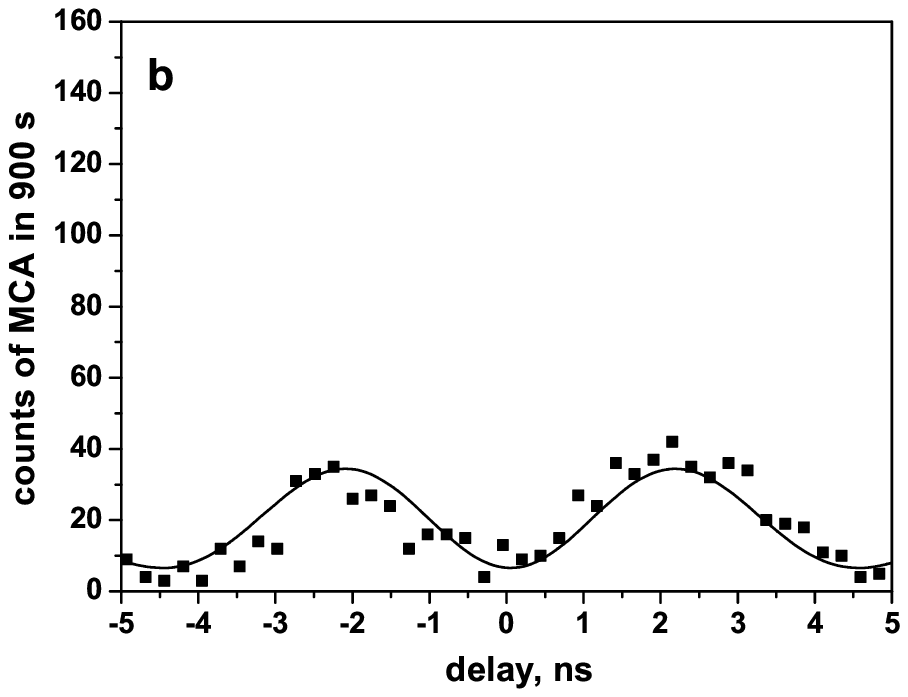}
\includegraphics[height=5cm]{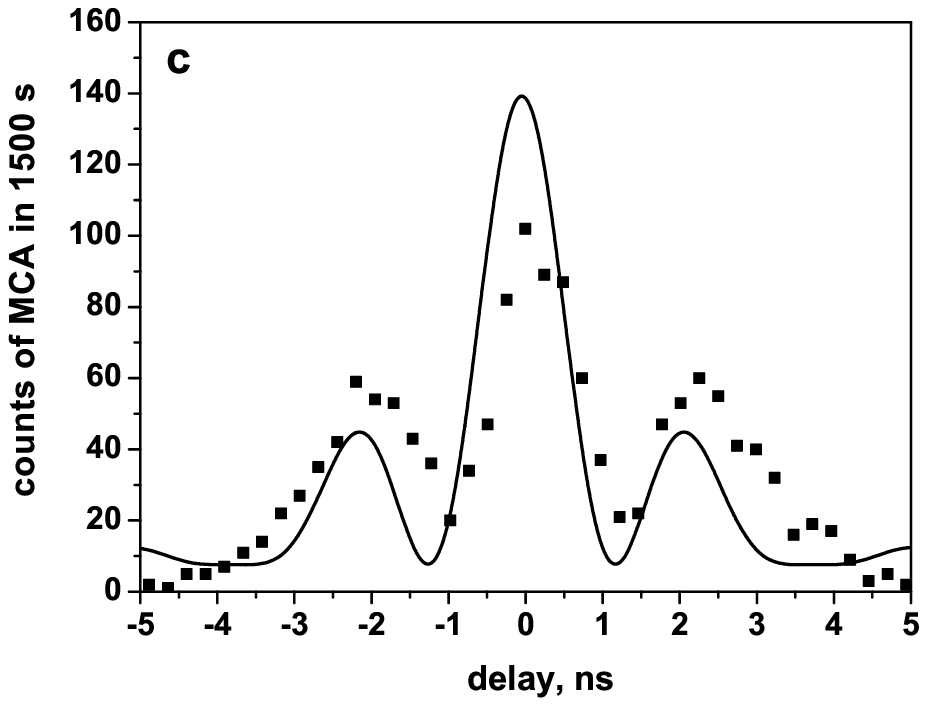}
\includegraphics[height=5cm]{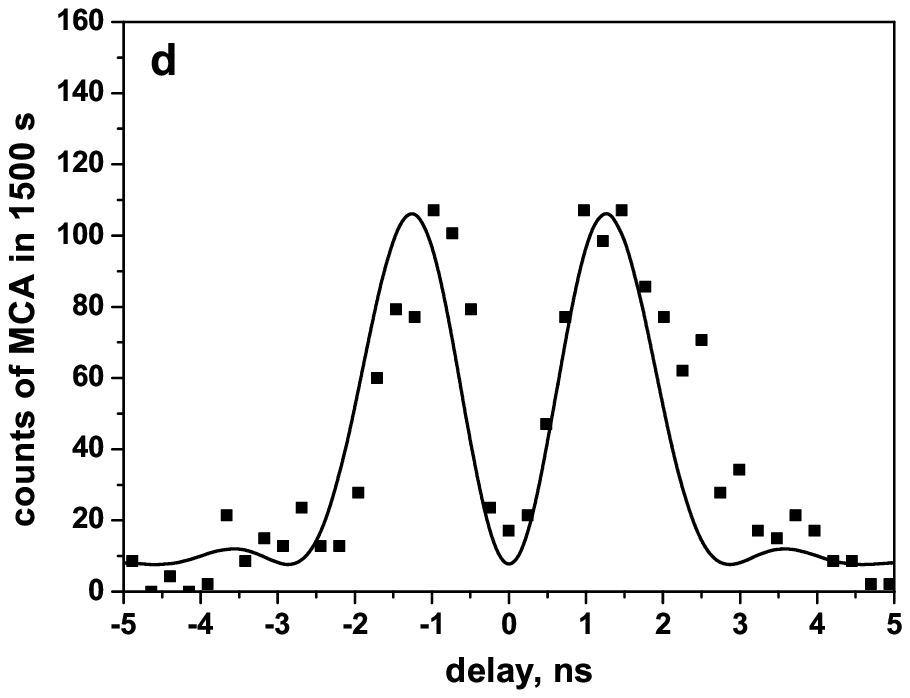}
 \caption{The measured shape of the second-order correlation function for the
case of a quartz plate of thickness 1 mm placed at the output of
the crystal with the optic axis parallel to the plane of the BBO
optic axis (a,b) and orthogonal to this plane (c,d), for two
different settings of the polarizers: $\theta_1=\theta_2=45^\circ$
(a,c) and $\theta_1=-45^\circ, \theta_2=45^\circ$(b,d). The length
of the fibre is 1 km. The curves show the theoretical dependencies
using ~(\ref{plates}).}
\end{figure}

\begin{figure}
\includegraphics[height=5cm]{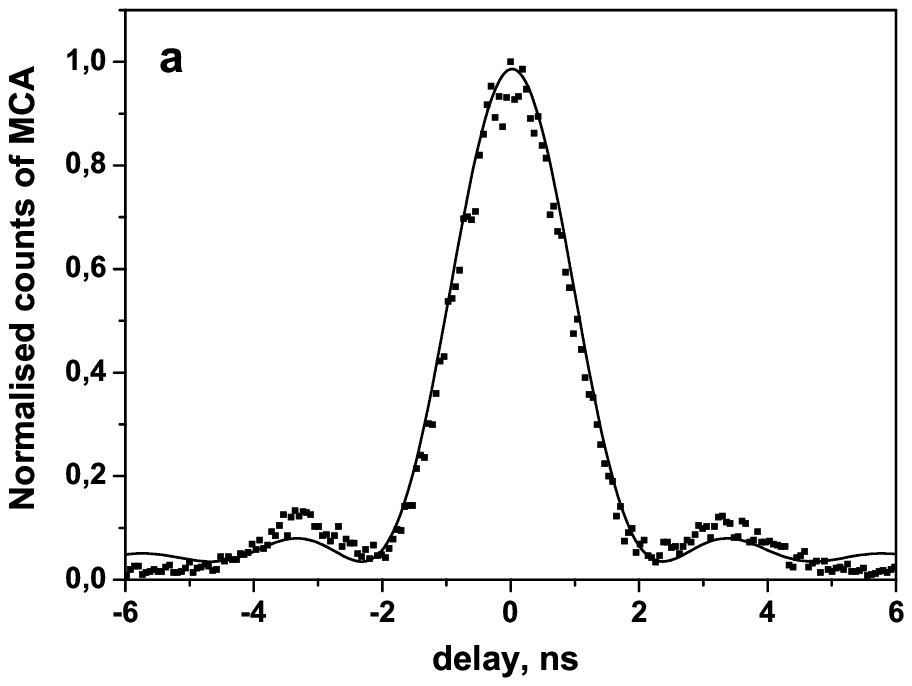}
\includegraphics[height=5cm]{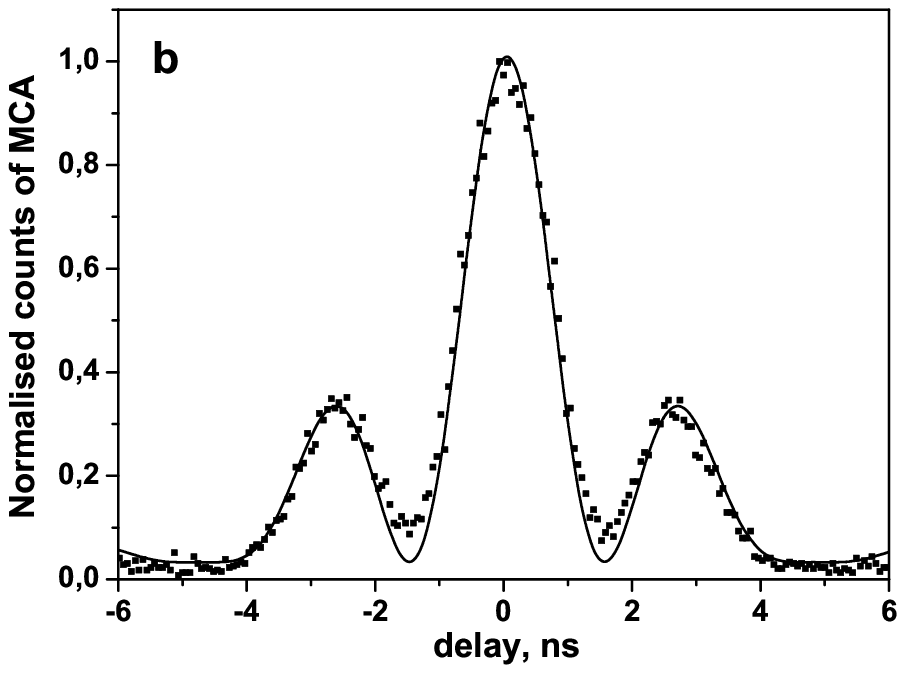}
\includegraphics[height=5cm]{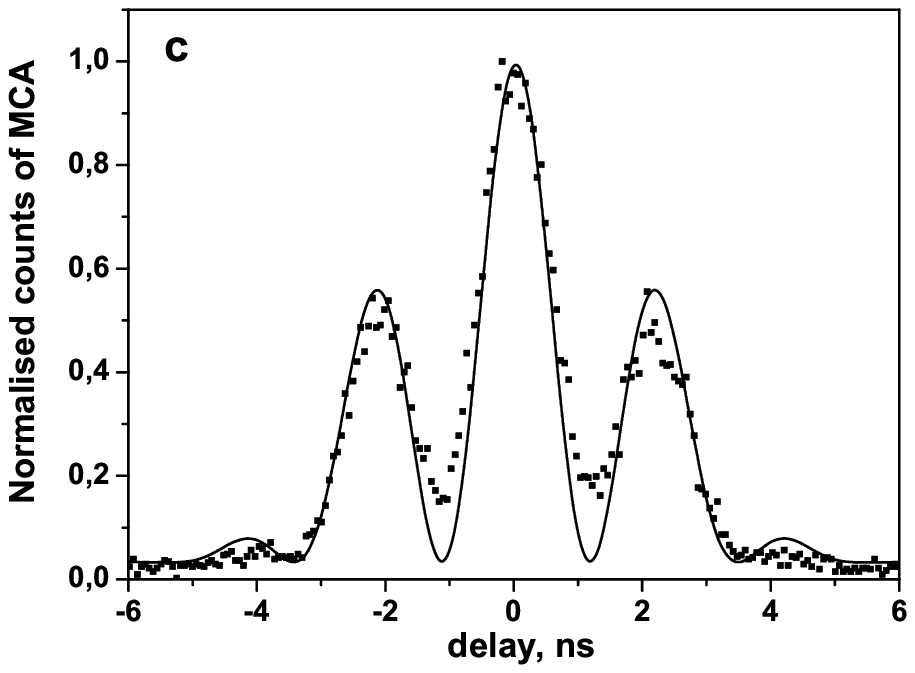}
 \caption{The measured shape of the second-order correlation function for the
case of (a) no quartz plates, (b) one quartz plate of thickness $1$
mm and (c) two quarts plates of  thickness $1$ mm placed at the
output of the crystal with the optic axes orthogonal to the plane of
the BBO optic axis (which increases the e-o delay), for the case of
a single polarizer introduced before the fibre. The polarizer is set
at $45^\circ$. The length of the fibre is $1$ km. The curves show
the theoretical dependencies using ~(\ref{plates}).}
\end{figure}


\begin{references}
\bibitem{distance} One can provide this by putting small pinholes at a certain
distance from the crystal where photon pairs are generated.
\bibitem{sufflength} A typical length of about 1 meter is
sufficient.
\end{references}
\end{document}